# Enhanced Hyperpolarized Chemical Shift Imaging Based on *a priori* Segmented Information

Gil Farkash, Stefan Markovic, Mihajlo Novakovic and Lucio Frydman*

*Department of Chemical and Biological Physics, Weizmann Institute of Science, Rehovot 76100, Israel*

# Abstract

**Purpose:** To develop an approach for improving the resolution and sensitivity of hyperpolarized $^{13}$C MRSI based on *a priori* anatomical information derived from featured, water-based $^1$H images.

**Methods:** A reconstruction algorithm exploiting $^1$H MRI for the redefinition of the $^{13}$C MRSI anatomies was developed, based on a modification of the Spectroscopy with Linear Algebraic Modeling (SLAM) principle. To enhance $^{13}$C spatial resolution and reduce spillover effects without compromising SNR, this model was extended by endowing it with a search allowing smooth variations in the $^{13}$C MR intensity within the targeted regions of interest.

**Results:** Experiments were performed *in vitro* on enzymatic solutions and *in vivo* on rodents, based on the administration of $^{13}$C-enriched hyperpolarized pyruvate and urea. The spectral images reconstructed for these substrates and from metabolic products based on predefined $^1$H anatomical compartments using the new algorithm, compared favorably with those arising from conventional Fourier-based analyses of the same data. The new approach also delivered reliable kinetic $^{13}$C results, for the kind of processes and timescales usually targeted by hyperpolarized MRSI.

**Conclusions:** A simple yet flexible strategy is introduced to boost the sensitivity and resolution provided by hyperpolarized $^{13}$C MRSI, based on readily available $^1$H MR information.

**Introduction**

Dissolution Dynamic Nuclear Polarization (dDNP) increases spin polarization by a factor of more than 10,000 (1), enabling *in vivo* $^{13}$C MR imaging (MRI) and spectroscopic imaging (MRSI) studies of labeled metabolites with good signal-to-noise ratio (SNR) despite their low concentrations (1, 2). dDNP hyperpolarizes a sample *ex situ*, by transferring an electron's spin polarization via microwave irradiation to the nuclear spins in the solid state, under cryogenic conditions and in a custom-made hyperpolarizer. Once the nuclear polarization has reached the desired level this solid sample is suddenly dissolved by flushing it out of the hyperpolarizer with a heated solvent, and rapidly transferred to a nearby MR scanner for injection into an animal or patient. Given a sufficiently long post-dissolution nuclear $T_1$, this allows one to track metabolism *in vivo* with unusual sensitivity. This is carried out by focusing on non-protonated $^{13}$C sites, which provide minute-long intervals until their polarizations return to undetectable, thermal equilibrium values. One of dDNP $^{13}$C MRSI's main uses has been to probe the glycolysis and TCA cycle, by injecting $^{13}$C$_1$-enriched pyruvate and monitoring the generation of $^{13}$C$_1$-lactate, $^{13}$C$_1$-alanine and $^{13}$C-bicarbonate (3, 4). This opens promising oncological perspectives, where hyperpolarized $^{13}$C$_1$-pyruvate and other substrates like fructose and glucose have been used to quantify the Warburg effect, thus localizing malignant tissue and evaluating treatment response (2–9). Additional *in vivo* applications of hyperpolarized MRSI have included cardiac imaging (10, 11), examination of muscle function (12), placental metabolism investigations (13) and the monitoring of perfusion – for instance by following the fate of $^{13}$C-urea in kidneys, tumors and other organs (14). The low toxicity of these metabolites, their stable isotope character and the lack of ionizing radiation, endows this MR approach with the prospect to evolve into a new form of metabolic molecular imaging, without some of the disadvantages associated to PET or CT.

A common demand of dDNP studies lies in their need to collect several independent pieces of information, within the limited time allotted by the lifetime of the nuclear hyperpolarization. These include spatial information in order to define the region/organ from which a given response is arising, as well as spectroscopic signatures characterizing the metabolic origin of a given response. MRSI can provide this multidimensional information in a number of different ways (15–19); for the present study we focused on a simple chemical shift imaging realization (20), whereby two spatial dimensions are independently scanned by centrically-sampled phase-encoding ($k_x,k_y$) variables, and a Free Induction Decay (FID) is subsequently collected as a function of a time $t$ encoding the chemical shifts. A main additional challenge arising in dDNP MRSI, particularly in metabolic-oriented experiments, stems from the need to collect these already high-dimensional data at multiple time points. This needs to be carried out at multiple instants, in order to track both the metabolic rates as well as the perfusion characteristics of the externally injected agent. When considering that the nuclear hyperpolarization will decay in ca. 60 seconds and that each scan will consume at least part of the available polarization, it follows that despite the ca. 10,000x signal enhancements brought about by dDNP, the MRSI sensitivity of this method is limited. To make up for this compromises in spatial resolution are usually taken, leading to voxel dimensions ranging from 100s to 1000s of µL. This may lead in turn to spectral contamination between spatially distal regions ('leakage'), limiting the clarity with which changes in a given anatomic region can be followed.

A number of routes allow one to improve spatial resolution without concomitantly increasing the number of scans. One is by relying on compressed sensing schemes (21–23), which accelerate MRI by randomly sampling the associated *k*-space, and then reconstructing the data using non-linear sparsity-based algorithms. Compressed sensing has been recently demonstrated for hyperpolarized $^{13}$C MRSI as a way to either reduce the number of excitations needed per image, and therefore reduce the consumption of hyperpolarization per completed image (24–26); it is also regularly used to reconstruct $^{13}$C images with higher spatial definition than what would be available if using uniform *k*-space sampling (24, 27, 28). Still, even with this aid, the fact that both the sensitivity and the image reconstruction demand a minimum number of values to be sampled close to the center of *k*-space, puts a limit to the overall acceleration that can be exercised in dDNP MRSI. An alternative way of reducing the number of phase encodes without losing spatial resolution, arises if assuming that metabolic signals are spatially uniform within certain physiological compartments. Consider for instance an MRSI scan based on the injection of a hyperpolarized substrate, where interests are focused on the substrate's uptake and metabolism within a number of *a priori* defined organs like kidney, liver, etc. If each compartment could be assumed to entail a nearly uniform spectral intensity this model would require very few phase-encoded acquisitions to deliver a highly defined, segmented information –in principle, no larger than the number of signal emitting compartments. Spectral Localization by Imaging (SLIM, 29), Spectral Localization with A linear Model (SLAM, 30) and Spectral LOcalization with an Optimal Point spread function (SLOOP, 31) make this regional uniformity assumption; it has been shown that all these methods can in principle reduce dramatically the number of scans required for MRSI, and in some instance provide an improved SNR/scan. A weakness of these methods rests in their disregard of regional non-uniformities in the $B_0/B_1$ homogeneities, but if these features are suitably mapped they can significantly improve the reconstructions (30, 32, 33). Additional errors will arise when the assumption of spectral uniformity within each compartment is invalid. To address this issue, more complex mathematical models that incorporate unknown distributions of metabolite concentrations within each compartment, have been proposed (34). This assumption is also lifted in this work, which explores the possibility of improving resolution and sensitivity in hyperpolarized $^{13}$C MRSI, by relying on a segmentation modality that does not assume spectral uniformity within compartments. To do so an approach is developed that assumes spectra to be locally akin in neighboring elements within a compartment, but still allows variability within their spectral intensities. Similar assumptions have been made for PET reconstructions that targeted spatial resolution improvement based on co-registered MRI or CT images, either by maximizing the correlation between edge maps of the functional and anatomical images (35–37), or by dividing the CT/MR images into a set of compartments through which PET pixels are similarly distributed (38). The former approach is analogous to widely used 'weak membrane' priors in optical imaging, in which an image is divided into smooth patches (39). From an operational standpoint our algorithm is slightly different from either strategy and we refer to it as PRO-SLAM, since it is based on both SLAM and on repeated projections of the data onto the sampled regions of *k*-space, of the kind used in the Projections Onto Convex Sets (POCS) partial Fourier reconstruction technique (40, 41). It is shown below that the ensuing reconstruction method can enhance the $^{13}$C MRSI's spatial resolution, improve SNR while maintaining spatial resolution, or provide the same image while reducing the number of *k*-samples and thereby preserve polarization for better kinetic characterizations. This study focuses in particular on the first of these options, with experimental comparisons between PRO-SLAM and MRSI acquisitions relying on Fourier transforming (FT) the same number of samples. Attention was also placed on the method's ability to deliver reliable kinetic characterizations. The following paragraphs describe in detail the flow of the proposed algorithm, and the experimentations done to validate its usefulness.

**Methods**

*Algorithmic Considerations.* 1D MRSI encodes its spatial information in the *k*-domain, and the spectrum as a function of a time-domain variable *t*. The resulting *s(k,t)* signal can be expressed as

$$s(k,t) = \iint \rho(r,f) e^{-i2\pi(kr+ft)} \, df \, dr \qquad (1)$$

As mentioned, PRO-SLAM aims to match the $^{13}$C MRSI outcome $\rho(r,f)$, to voxels whose shapes have been predefined by an anatomical $^1$H MRI scan. To do this we start, as in SLAM (30), by subjecting the spectral domain to a conventional $FT_t$ that defines the spectral frequencies; we can then focus purely on imaging reconstruction aspects for each peak in the resulting *f*-domain. Assuming for simplicity a single spatial dimension and taking into account the discrete sampling of all MR variables, (1 can be rewritten for each specific *f* as:

$$\begin{bmatrix} s(k_1) \\ s(k_2) \\ \vdots \\ s(k_M) \end{bmatrix} = \begin{bmatrix} e^{-i2\pi k_1 x_1/M} & e^{-i2\pi k_1 x_2/M} & \cdots & e^{-i2\pi k_1 x_M/M} \\ e^{-i2\pi k_2 x_1/M} & e^{-i2\pi k_2 x_2/M} & \cdots & e^{-i2\pi k_2 x_M/M} \\ \vdots & \vdots & \ddots & \vdots \\ e^{-i2\pi k_M x_1/M} & e^{-i2\pi k_M x_2/M} & \cdots & e^{-i2\pi k_M x_M/M} \end{bmatrix} \times \begin{bmatrix} \rho(x_1) \\ \rho(x_2) \\ \vdots \\ \rho(x_M) \end{bmatrix} \qquad (2)$$

This matrix form can be expressed as $s_M = PE_{M \times M} \times \rho_M$, where the phase-encoding *PE* matrix is for all practical purposes an FT operator, whose inversion yields the sought intensities $\rho$. SLAM assumes *a priori* that intensities are grouped into uniform regions within *C* compartments, making (2 over-determined. This can be exploited to reduce the number of sampled elements; using the notation just mentioned, this would imply that $s_{M'} = PE_{M' \times C} \times \rho_C$. As *C* is usually smaller than *M*, this in turn means that it becomes possible to reconstruct the imaging information with a correspondingly smaller number of samples *M'*.

A more flexible assumption about the image intensity inside a compartment will result if the highly defined boundaries of the compartments are respected, but the available *k*-space data is employed to define the intra-compartment distributions in a less strict way –for instance by requesting that elements are spatially similar rather than identical. A reconstruction algorithm that assumes such spatial similarity but allows for smooth variations was developed for POCS, an iterative procedure used to reconstruct high-resolution, purely-absorptive images despite missing *k*-space data (55-57). For a given $k_{max}$ sampling support, POCS ultimate goal is to estimates data missing in the $-k_{max} \leq k \leq 0$ interval from data sampled in $0 \leq k \leq k_{max}$, by exploiting the real character of the images. This is done using an iterative algorithm that estimates the missing signal while preventing rapid local phase changes in the images –a constraint that has been shown equivalent to a brute force matrix inversion leading from the available data to the desired image, while being faithful to a minimization of the $L^2$ norm (56-59). Similarly, the PRO-SLAM algorithm hereby proposed uses iterations to complete the missing *k*-space data carrying the high-resolution features, while enforcing intra-compartment smoothness in the image domain by minimizing amplitude gradients within each compartment. More specifically, let's assume that to accommodate the high definition that would be desired in $^{13}$C MRSI one would have to collect data over $-k_{max} \leq k \leq +k_{max}$, but that for practical reasons data were only sampled for a central portion of *k*-space $-k_{PRO\_SLAM} \leq k \leq +k_{PRO\_SLAM}$, with $k_{PRO\_SLAM} \leq k_{max}$. The missing high frequency samples in the $|k_{PRO\_SLAM}| \leq |k| \leq |k_{max}|$ interval define a convex set, in the sense that any linear combination of two images that respects the recorded central *k*-space data and fulfills intra-compartment smoothness, will also be an image that belongs to this set. Hence, recovering the missing *k* samples is a problem that can be solved by an iterative procedure of the kind summarized in Figure 1; for each spectral frequency element *f,* this algorithm operates as follows:

(1) Low spatial resolution $^{13}$C images are reconstructed by the inverse FT of zero-filled *k*-space data, where the original $-k_{PRO\_SLAM} \leq k \leq +k_{PRO\_SLAM}$ data is augmented to the range $-k_{max} \leq k \leq +k_{max}$ required to satisfy the resolution defining the $^1$H images to be segmented

(2) Smoothness within the segmented high-resolution $^1$H image is enforced on the $^{13}$C data by replacing each pixel in the segmented compartment, with a weighted average of their neighbors. For 2D image acquisitions of the kind carried out in this study, this entails the replacement

$$\rho_{new}(x_0, y_0) = \frac{\sum_{n=1}^{N} w_n \rho(x_n, y_n)}{\sum_{n=1}^{N} w_n} \qquad \forall (x_0, y_0) \in Compartment_m \qquad (3)$$

where $\rho(x_n, y_n)$ are the image elements within the compartment, and $w_n$ denotes a Gaussian weighting depending on the relative distance of any of these elements with respect to $(x_0, y_0)$. Equation 3 can be seen as a Gaussian filter convoluted with a binary mask set to 1 for a compartment member and zero otherwise, so as to preserve sharp compartment edges. The full-width-half-maximum associated with the Gaussian weights $w_n$ is a user-defined parameter controlling the potency of this regularization; if this is set very wide (i.e., constant weighting throughout the chosen compartment) this algorithm would revert back to the regular SLAM regime.

(3) FT is applied on these smoothed images to calculate a "corrected" set of *k*-space data.

(4) The low frequencies of this new *k*-space data set are then reset to their original, experimentally measured values, while the data that had been originally zero-filled has now gained non-zero elements. These can be left as such, or the transition around $k_{PRO\_SLAM}$ can be "tapered" to ensure consistency between the acquired $-k_{PRO\_SLAM} \leq k \leq +k_{PRO\_SLAM}$ central data and the estimated high-frequency $|k_{PRO\_SLAM}| \leq |k| \leq |k_{max}|$ components, thereby prevent ringing artifacts.

(5) A new set of $^{13}$C images is estimated by FT of these new data

(6) Steps 2-5 are iteratively repeated until the $L_2$ norm of differences between the new *k*-space data and the one calculated in the previous iteration are small.

(7) Once the final iteration is done, inverse FT of the augmented *k*-space data delivers the PRO-SLAM image.

The algorithm just described defines a search within a complex set: in other words, if the segmented compartments are common and these have been smoothed by a procedure that in all instances respects the central *k*-space points that have been collected, then any linear combination of two images belonging to this set will also exhibit smooth compartments and retain the central *k*-space points that were sampled. Indeed, consider two images $\rho_1$ and $\rho_2$ belonging to a common set that respects the recorded central *k*-space data and fulfills intra-compartmental smoothness; then, a convex combination of $\rho_1$ and $\rho_2$ can be written as $\rho_3 = s\rho_1 + (1-s)\rho_2$, where *s* is a constant. To visualize why this is so we recall that, due to the linearity of the FFT, the low frequency *k*-space associated with $\rho_3$ will also be consistent with the recorded low frequency *k*-values. As for $\rho_3$'s fulfillment of inter-compartmental smoothness –the algorithm above defines this as an element-wise multiplication of a low-pass filter kernel *g* -chosen above to be a Gaussian- times a binary mask *b* that spatially selects compartment members, both extending over *N* points. Denoting these weighting elements by the shorthand $w = g \cdot b$, one can summarize the smoothed images as $\rho_1 = conv(v_1, w)$ and $\rho_2 = conv(v_2, w)$, where the *v*'s represent unfiltered images. Expressing now these convolutions as one-dimensional discrete sums and focusing on a specific pixel $r_o$ leads to

$$\rho_3(r_o) = s\sum_{m=-\frac{N}{2}}^{\frac{N}{2}} v_1(r_o - m)w(m) + (1-s)\sum_{m=-\frac{N}{2}}^{\frac{N}{2}} v_2(r_o - m)w(m) \qquad (4a)$$

$$= \sum_{m=-\frac{N}{2}}^{\frac{N}{2}} [sv_1(r_o - m) + v_2(r_o - m) - sv_2(r_o - m)]w(m)$$

$$= conv[sv_1(r_o) + (1-s)v_2(r_o), w] \qquad (4b)$$

which is also an image satisfying intra-compartmental smoothness. This proof of the convexity criteria can be generalized further to cases where compartments have different spatial variabilities, by defining the cutoff of the applied low pass filter as given by the maximal frequency of these variations.

*Experimental.* The PRO-SLAM reconstruction just described was compared against FT-based MRSI results in a series of thermal $^1$H, thermal $^{13}$C and hyperpolarized $^{13}$C *in vitro* and *in vivo* experiments. All $^{13}$C data were acquired on a 4.7 T Bruker (Ettlingen, Germany) Biospec® horizontal scanner running Paravision® 4.1 and controlled by an Avance® console. The setup included a double-resonance, crossed-coil configuration, transmitting using a $^{13}$C Bruker volume coil and receiving through a 20 mm Doty Scientific surface coil (Columbia, South Carolina) while applying active decoupling. In some of the experiments, hyperpolarized MRSI was used to monitor $^{13}$C$_1$-pyruvate metabolism, while in others perfusion was tracked on hyperpolarized $^{13}$C urea. Anatomical $^1$H images were co-recorded with these experiments using either the surface or volume coil, with a gated FLASH sequence. These images were acquired with 6 cm in plane fields-of-view (FOV), and echo/repetition times TE / TR of 6.3 / 615 ms respectively.

$^{13}$C$_1$-pyruvic acid and $^{13}$C urea (Cortecnet, Voisins Le Bretonneux, France) were polarized with 15 mM Ox63 trityl radical (GE Healthcare, Little Chalfont, UK) in an Oxford Instruments Hypersense® (Abingdon, UK) polarizer, by applying microwave irradiation at 94.1 GHz while the sample was placed at 1.3 K inside a 3.35 T magnet. Following the polarization stage, the solid sample was dissolved with either superheated Tris buffer (for pyruvate) or PBS (for urea), thereby bringing it to physiological (30-35 ˚C) temperatures. These liquid samples were transferred ca. 3 m into a syringe placed nearby the MRI, for injection and subsequent acquisition of the hyperpolarized MRSI scans.

While the main focus of this study is on $^{13}$C acquisitions, it was easier to quantify the performance of the PRO-SLAM approach on $^1$H MRSI experiments, as these are endowed with much higher sensitivity and with the possibility of rerunning multiple tests under different acquisition conditions. These experiments were carried out at 7 T using a Millipede® linearly polarized volume probe on a VNMRS® 300/89 vertical bore microimaging Varian system, using an axially symmetric phantom. The pulse sequence employed in such experiments had a non-selective pulse, followed by two adiabatic band selective LASER pulses that refocused spins over a 4 mm slice perpendicular to the axis of symmetry. *k* data were collected at three different in-plane spatial resolutions –16x16, 32x32, 64x64– lasting 15 minutes, an hour and 4 hours acquisition times, respectively. PRO-SLAM reconstructed all spectral images at a target resolution of 128x128. These data were employed to quantify correlation to the high definition reference image by measuring normalized distances of each pixel from its analogue in the reference image: $\Delta_{ref} = 100 \cdot \frac{|s(x,y) - s_{ref}(x,y)|}{|s_{ref}(x,y)|}$ and then averaging those distances in the relevant ROI. FT images used in these $\Delta_{ref}$ comparisons were interpolated to the same spatial resolution as the PRO-SLAM images (128x128).

All MRSI experiments relied on a home-written center-out *k*-sampled chemical shift imaging sequence (15), incorporating an initial small-flip-angle Gaussian pulse that selected a *z*-slice of 5 mm for the urea or 10 mm for the pyruvic acid experiments. These pulses were set to 25˚ nutations for the urea experiments; concurrent 25˚/5˚ nutations were used for exciting the lactate/pyruvate resonances in the pyruvic acid injection experiments. These values were selected for maximizing SNR in the kinetic analyses. $^{13}$C MRSI acquisitions involved repetition times TR of 68 ms for the $^{13}$C pyruvate and 46 ms for the $^{13}$C urea experiments. To encode the imaging information, 12x12, 10x10 or 8x8 phase encoded FID's were collected in $k_x$-$k_y$, leading to overall acquisition times of 8.3, 5 or 3 sec, respectively. For both pyruvate and urea the scanned Fields-of-View (FOVs) were 5x5 cm$^2$, and 256 time points were collected with a bandwidth of 5.9 kHz (119 ppm) along the spectral dimension. In general, 10 or 11 MRSI data sets could be acquired by repeating these pulse sequences in the same center-out *k*- trajectory, before the hyperpolarized signals decayed into the noise.

Three kinds of *in vitro* experiments were conducted. In the first set of test, $^1$H MRSI scans were reconstructed from a phantom containing an acetone Eppendorf immersed in a water tube. In addition, thermal $^{13}$C MRSI data were collected on a tube filled with enriched $^{13}$C$_1$-glucose (25%) dissolved in water, placed adjacent to a tube containing only water. The ensuing $^1$H-coupled $^{13}$C spectrum consisted of two overlapping doublets, leading to a triplet-like appearance. Accompanying $^1$H FLASH images were also recorded to define the compartments needed as input for the PRO-SLAM reconstruction. In order to compare SNR measurements of images arising from PRO-SLAM against their conventional FT counterparts, all spectroscopic images were zero filled to the same size (42x42 in plane points); SNR was then measured for each *x-y* coordinate by dividing the mean peak intensity by the root-mean-square of 50 sample points positioned in the noise. An additional *in vitro* experiment was performed to monitor the quantitative reliability of PRO-SLAM vs FT $^{13}$C MRSI, using a calibrated kinetic system. To this end hyperpolarized pyruvic acid was injected into an Eppendorf tube containing lactate dehydrogenase (LDH), an enzyme that catalyzes production of lactate using pyruvate as substrate (46). The assay mixture containing the enzyme had a neutral pH and included 80 mM of the co-activator NADH (Bio-world, Dublin, Ohio) in Tris buffer, and 240 units/mL of LDH (SigmaAldrich) originating from rabbit muscle (47). Hyperpolarized pyruvate was also injected into an adjacent Eppendorf tube which contained only phosphate buffer for control. A third tube containing water was placed in the middle of these two Eppendorf tubes for shimming purposes. Following the pyruvate's hyperpolarization, 800 µL of 80 mM $^{13}$C$_1$-pyruvate were injected in equal aliquotes into the right- and left-hand side Eppendorf tubes, and a time-series of 15 spectral images was collected over approximately 1.5 minutes.

*In vivo* $^{13}$C hyperpolarized MRSI experiments were also carried out, focusing on placentas of Wistar pregnant rats. Nine rats were scanned at days 17-21 of gestation, with pregnancy timed on the morning following overnight pairing of male/female pairs. Six of these rats were injected via their tail veins with 80 mM of hyperpolarized $^{13}$C$_1$-pyruvate, and three rats were injected with 115 mM of hyperpolarized $^{13}$C-urea. In all cases the injection bolus was 3 mL, and rats were anesthetized by inhaling a mixture of 3% isoflurane and oxygen fed for respiration through a dedicated nose mask at 1L/min. Images and spectra were collected on these animals by placing the surface coil on the abdomen of the rats. These experiments were approved by the Institutional Animal Care and Use Committee of the Weizmann Institute of Science (IACUC App #36350617), and some of these data has been presented and its biological content analyzed in Ref. 13.

The PRO-SLAM reconstruction algorithm was executed using Matlab (Natick, MA), based on compartments that were defined over $^1$H structural images. For the *in vivo* scans these compartments were manually drawn; for the *in vitro* phantom tests, endowed with better signal-to-noise, compartments were automatically segmented using a simple thresholding of $^1$H reference images. In all $^{13}$C MRSI cases the datasets were first subject to time-domain apodization using a 20-40 Hz exponent, inverse FTd vs *t,* baseline corrected, and onward image-processed for each desired spectral element. For PRO-SLAM the in plane image matrices were 42x42 for all *in vivo* experiments, while for the FT processing datasets were zero filled to produce 18x18 or 20x20 image matrices. For the thermal glucose phantom case, the PRO-SLAM and FT images were zero-filled to 42x42 points for a better SNR comparison; for the 1H MRSI *in vitro* tests described in the SI all images were reconstructed using 128x128 in-plane matrices. The Gaussian weights $w_n$ used by the regularization enforcing compartmental smoothness were calculated based on a variance $\sigma^2 = 1$ or 4 pixels for all compartments, except for the background. The background was considered as an additional compartment and smoothed more aggressively in the dynamic MRSI studies ($\sigma^2 = 4$) pixels. For the thermal glucose phantom experiments, the background compartment $\sigma^2$ was set to 1 like for the other compartments. These $\sigma$ values of the Gaussian weights were designed to preserve the spatial variations in metabolite signal –coming from $B_0/B_1$ inhomogeneities or concentration variations– while attenuating most of the noise at higher frequencies. Even though the algorithm had a concrete stopping criterion based on tracking changes between iterations, in practice images required ≤5 iterations before reaching convergence. The typical PRO-SLAM processing required 30 seconds per metabolite spectral peak, when running on a desktop PC with an Intel Core i7-3770 processor. Thus an overall processing time of one minute was required for processing an *in vivo* hyperpolarized $^{13}$C pyruvate experiment, accounting for both the pyruvate and lactate spectral bins.

**Results**

Figure 2 compares *in vitro* thermal data reconstructed, for a high sensitivity $^1$H phantom made out of coaxial water and acetone tubes. Shown in panels (a,b) are the images reconstructed for various acquisition matrices by FT and PRO-SLAM for the water resonance; panels (c,d) show the results of the reconstructions for the acetone peak bin. The improved uniformity, sharpness and SNR afforded by PRO-SLAM can be appreciated from these images, particularly for cases involving constrained k-space acquisitions. Further quantitative aspects of these tests, including image correlations vs the reference image, the image contrast arising for both methods and quantitative reliability issues, are summarized in Figure S1 and Tables S1 and S2 of the Supporting Information section.

Figure 3 extends these measurements to $^{13}$C MRSI tests, based on a $^{13}$C$_1$-enriched glucose sample. These reconstructions arise from the same data and focus on two spectral regions: one containing the glucose peaks, and one containing noise. As can be seen form the noise maps in the right column, noise is significantly reduced by PRO-SLAM: for the glucose-containing spectral region, the average noise intensity produced by FT was twice as high as that arising in PRO-SLAM. The image intensities originating from the glucose peaks are similar for both processing protocols, but become more uniform for PRO-SLAM within the sample tube. Also the image's edges are sharper, and spill-overs outside the sample tube are reduced. Residual heterogeneities still remain within the sample tube, probably reflecting the sensitivity profile of the surface coil used in this experiment. An artifact in the form of a delta function appears in the center of *k* space for both PRO-SLAM and FT; arising to a residual DC bias on the time domain signals.

Figure 4 displays a time series of *in vitro* hyperpolarized experiments, aimed at exploring the quantitative reliability of PRO-SLAM to follow chemical kinetics. Towards this end LDH, the enzyme that catalyzes the conversion of pyruvate to lactate, was placed in one tube, a solution with PBS (buffer) was placed in another, and the MRSI results arising from a simultaneous injection of hyperpolarized $^{13}C_1$-pyruvate into both tubes was followed. Spectral dimensions were processed, and the images corresponding to the injected pyruvic and to the emerging lactate peaks were reconstructed with FT and PRO-SLAM. For both methods, signals from pyruvate, lactate and pyruvate hydrate (the latter not shown) were observed up to the fifth experimental repetition, by which time the signals were lost in the noise. Both processing methods showed strong pyruvate and lactate signals in the tube containing the LDH, even if with the PRO-SLAM processing signals are better localized within the tube perimeter. Both methods also show a stronger, longer lasting pyruvate signal in the tube lacking the LDH, as in this environment the chemical is not consumed. In addition, both experiments give a weak artificial lactate response in the LDH-free tube –most likely spectral residues coming from broad pyruvate peaks whose wings appear at lactate's chemical shift due to $B_0$ inhomogeneities. Most importantly, this experiment confirms that PRO-SLAM and FT yield similar kinetics, thereby enabling one to rely on the former's quantitativeness.

PRO-SLAM methods were also used to evaluate the uptake and metabolism of pyruvate in pregnant rats. Recent work has demonstrated the feasibility of using hyperpolarized pyruvate to noninvasively examine fetoplacental transport in guinea pigs and chinchillas (48, 49); it has also been used to explore how different metabolites behave in the placentas of naïve and diseased pregnant rats (13). These studies are challenging as they have to deal with small spatial entities liable to spill-over effects, as well as with limited sensitivity. Figure 5 shows representative $^{13}C$ hyperpolarized MRSI outcomes, of *in vivo* studies exploring the conversion of pyruvate into lactate in healthy pregnant rats. In these images pyruvate signals emanate mainly from placentas, and monotonically decline from the first image onwards. Lactate $^{13}$C signals by contrast rise gradually from the beginning of the time-series, peak in the fifth/sixth frame, and thereafter decline over the rest of the time series. Overall, the PRO-SLAM data show a more compact localization of the pyruvate signal in the placentas marked as input for the algorithm; also the lactate PRO-SLAM results show a tighter matching to the anatomy. This is notable for the two closely placed placentas (P1, P2) in Figure 5: the FT-derived images show a spillover at their interface, both for the initial pyruvate and the middle lactate frames in the time series. In PRO-SLAM by contrast, the metabolites are spatially divided into two compartments with clear boundaries. A superior PRO-SLAM localization is also shown in Supporting Information for five additional hyperpolarized MRSI studies, which also monitored the metabolism of $^{13}$C$_1$-pyruvate and the conversion to lactate. Similar metabolite dynamics are once again seen in all these studies for both PRO-SLAM and FT (Figure 6 and Supporting Figure S3), with minor differences stemming from closely-spaced organs liable to spatial spill-overs.

Figure 7 shows results from the MRSI study reported as #2 in Supporting Information Figure S2, only this time focusing on the spectral range corresponding to the $^{13}$C$_1$-alanine being metabolically produced upon injecting the pyruvate. The $^{13}$C$_1$-alanine peak was prevalent for this slice in two placentas, in a fetal liver and in a maternal kidney. Images for the alanine chemical shift peak were accordingly reconstructed by PRO-SLAM with four compartments as input, and compared to their FT counterparts. According to both processing methods the alanine signal from the maternal kidney and the fetal liver decayed continuously throughout the experiment; by contrast, for the placental compartments, both methods showed a signal that peaked soon after the tail vein injection, and then decayed monotonically.

Figure 8 illustrates results recorded upon the *in vivo* infusion of hyperpolarized $^{13}$C-urea into a pregnant rat, in a slice targeting mainly maternal organs. The spatial resolution is better for the images in these urea experiments than for the pyruvate counterparts above, thanks to the use of a thinner slice thickness (5 mm for urea; 10 mm for pyruvate). The maternal kidneys and the maternal vena cava are better edged out in PRO-SLAM, as the spatial leakage of the urea signal is reduced. In this study the kinetics of the perfused urea revealed by the processing algorithms were similar for one of the maternal kidneys, but differences arose in the curves of the second kidney and of the vena cava. We ascribe these differences to the significant spill-over that characterizes FT in this particular case. Supporting Figure S4 shows two additional cases of urea injection. Figure S4a includes five compartments that emitted significant $^{13}$C signals in this study, arising from a different slice of the same animal as shown in Figure 8: these were a maternal kidney, a fetal liver and three placentas. In all three placentas, PRO-SLAM and FT showed an increase until the second time-frame and then a constant decline. Perfusion dynamics were also similar for the maternal kidney and fetal liver. Notably, there was a significant sensitivity gain in PRO-SLAM for certain pixels (e.g., 'P2' in Figure S4a). In addition, PRO-SLAM delivered higher spatial definition near the compartments that were used for the reconstruction and better matching to the structural image. Figure S4b shows another example, where signals arising from the maternal kidney show similar kinetic curves by FT and PRO-SLAM.

**Discussion and Conclusions**

Linear analyses were initially considered as a way of restoring to FT-based MRSI, a spatial resolution which sensitivity could not support. For this, models were developed whereby spectral uniformity would be enforced within the compartments arising from sensitive, water-based $^1$H MRI measurements (29–31). Later developments such as G-SLIM highlighted the importance of modeling non-idealities (in fields, composition and concentrations) that may violate this uniformity (34); but modeling and solving for these intra-compartmental heterogeneities can be computational complex, and lead to numerical instabilities in the matrix inversion process that will result in unpredictable spectral localization errors. Still, these developments could have valuable uses in the equally challenged scenario of hyperpolarized spectroscopy; in this instance, however, additional validations are needed given the limited acquisition times that this method can support, and the reliance of hyperpolarized studies on the quantitation of a time-dependent kinetic signal. A recent study for instance showed the benefits of incorporating a pre-defined kinetic metabolic model as well as zeroing signals from outside an animal's body, when setting up a constrained reconstruction algorithm capable of improving the hyperpolarized $^{13}$C MRSI results (50). The use of 2D spatially selective RF pulses was also considered to excite certain compartments individually and record only their signal (51), yet the long RF pulses that this demands makes such approach susceptible to B$_0$ inhomogeneities, and T$_2$–driven SNR losses. The present paper extends these efforts by introducing PRO-SLAM, a reconstruction algorithm that fulfills these criteria by relying on predefined segments and enforcing smoothness –but not uniformity– within regions. The algorithm was devised based on an iterative convex reconstruction that completes non-sampled elements in *k*-space, while preserving fidelity to the portion of *k*-space that was physically sampled. A regularization that enforces slow spatial variability in the image domain is also present, controlled by Gaussian weights that replace each pixel's signal with a weighted average of its nearby elements. To investigate the effectiveness of this approach, an array of *in vitro* and *in vivo* conditions were explored. *In vitro,* $^{13}$C and $^{1}$H MRSI experiments confirmed that PRO-SLAM can improve SNR, edge contrast and spatial resolution. These control experiments showed that noise could be reduced by PRO-SLAM by factors of 2-3, while leading to images with sharper edges and definitions akin to the protons images. In addition, *in vitro* hyperpolarized MRSI studies monitoring the enzymatic turnover of $^{13}$C$_1$-pyruvate onto $^{13}$C$_1$-lactate showed that PRO-SLAM also maintains kinetic faithfulness, producing dynamic curves that are akin to the FT-based results. These quantitative reliability measurements were also backed up by thermal $^{1}$H MRSI determinations (Supporting Table S2). For *in vivo* testing attention was turned to hyperpolarized $^{13}$C MRSI on the relatively challenging case of abdominal scans in pregnant rodents. Also these studies confirmed PRO-SLAM's SNR and spatial definition advantages, while yielding reliable pyruvate→lactate and pyruvate→alanine conversion kinetics. $^{13}$C-urea studies highlighted similar perfusions into the maternal and fetoplacental organs, even if in some *in vivo* cases the metabolite kinetics observed didn't coincide. These cases, however, were associated to spillover effects in the FT-based images, which could be conducive to quantification artifacts that were alleviated by the PRO-SLAM protocol. Also in need of further research are the ways in which this algorithm's improvements depend on the experimental SNR, and on the rate at which the excitation pulses burn the relaxation-limited lifetimes of these hyperpolarized experiments.

    A number of extensions could further improve the algorithm here presented. One of these would be the incorporation of $B_0$ inhomogeneity effects that can be rapidly estimated by $^{1}$H MRI; these could be accounted in the algorithm by replacing the FT in step 1 of Figure 1 with a forward projecting model-based transformation that accounts for $B_0$-induced dephasings, and by a corresponding inversion matrix in step 4 (32, 33, 52). Likewise, RF inhomogeneities could be taken into account by modeling sensitivity spatial profiles as part of the aforementioned matrices – even if these may be harder to come by given the non-coincidence of $^{1}$H and $^{13}$C $B_1$ profiles even in double-resonance surface coil assemblies. Sparse sampling principles (36-42, 64) could also be

incorporated into the PRO-SLAM algorithm, probably at little cost performance-wise. Additional developments could include automatic detection of compartments for *in vivo* datasets based on edge-contrast, geometry or texture, of the kind that were here proven successful in high sensitivity *in vitro* experiments. The need to define –either manually or automatically– metabolically relevant compartments based on $^1$H anatomical images, may constitute a limiting aspects of the PRO-SLAM procedure: defining misleading compartments may result in degraded –rather than improved– $^{13}$C images, as relevant pixels inside these compartments may end up convoluted with neighbors carrying little or no information. This scenario highlights the importance of making a physiology-based decision upon defining compartments. Alternatively, if this information is unavailable, lowering the Gaussian weights used may reduce the presence of artifacts, at the expense of PRO-SLAM's SNR. It is worth mentioning that similar information/resolution challenges are being actively tackled in other MRI-related hybrid imaging modalities, by use of machine-learning tools (53–55). Several of the extensions mentioned, as well as tests in additional *in vivo* scenarios, are currently being explored.

**Acknowledgments.** We are grateful to Dr. Nava Nevo (WIS) for initial help in preparing and handling the animals, to Dr. Tangi Roussel (Neurospin) for Paravision programming assistance, and to Drs. Gilad Liberman (WIS) and Yi Zhang (Johns Hopkins) for valuable discussions about the algorithm. This research was supported by Minerva Project 712277, NIH Grant R01HD086323, the Kimmel Institute for Magnetic Resonance (WIS), and the generosity of the Perlman Family Foundation.

**Supporting Information.** Additional information can be found in the Supporting Information section available in the on-line version of this article.

# FIGURE LEGENDS

**Figure 1:** Schematic diagram of the $^{13}$C MRSI PRO-SLAM reconstruction algorithm implemented to achieve $^1$H-derived compartmental definition while allowing for smooth intra-compartmental variations within the hyperpolarized images. The inputs consist of the $^{13}$C MRSI *k*-space data for any given spectral frequency, and segmented compartments representing physiological relevant regions deriving from a high-definition anatomical $^1$H images (top, bold). The output is the $^{13}$C image (bottom, bold).

**Figure 2:** Comparison between $^1$H MRSI scans of a phantom reconstructed with PRO-SLAM and with FT. All data were acquired at 7 T using a volume RF probe and the phantom depicted in (f) containing an acetone Eppendorf immersed in a water tube. (a) Images corresponding to the water spectral window reconstructed after FTing in all three $(k_x, k_y, k_f)$ dimensions, using $k_x/k_y$ matrix sizes of 16x16 (left column) 32x32 (middle column) and 64x64 (right column). (b) Idem but processed with PRO-SLAM, and seeking a spatial resolution of 128x128 for all input datasets. The algorithm received as input two compartments that were automatically segmented from the structural LASER image in (e), corresponding to the acetone and water compartments. (c,d) Idem as (a,b) but upon reconstructing the acetone spectral bin. See Supporting Information for additional quantitative comparisons.

**Figure 3:** Noise, signal and sensitivity analyses arising upon processing by FT (a) or PRO-SLAM (b) spectrally-binned images of a thermal, $^{13}$C$_1$-enriched glucose sample yielding the spectrum on the left. Spatial images are displayed on the middle column in a common intensity scale, averaging the area marked in blue over the spectra in the left column. The maps on the right column represent the noise measured on the spectral region marked in red for the same x-y coordinates. (c) Ratio between the SNRs afforded by PRO-SLAM (P-S) and FT. A single compartment was defined for the PRO-SLAM reconstruction highlighted by a green contour on the $^1$H image (d). The setup also included a glucose-free water tube adjacent to the $^{13}$C-labeled solution tube for shimming purposes.

**Figure 4:** *In vitro* experiments exploring the PRO-SLAM's (lower row in each panel) ability to yield a kinetics that is comparable to FT-based MRSI (upper row in each panel). Images extracted at (a) the $^{13}$C$_1$-pyruvate and (b) the $^{13}$C$_1$-lactate chemical shifts, monitored during the enzymatic conversion of $^{13}$C$_1$-pyruvic acid into $^{13}$C$_1$-lactate. The setup (d) involved an Eppendorf containing LDH catalyzing the production of lactate from pyruvate, another without LDH acting as reference,

and a center one for $^1$H shimming and calibrations. Hyperpolarized $^{13}$C$_1$-pyruvate was simultaneously injected in the right and left containers; the resulting $^{13}$C MR images are displayed overlaid on top of their $^1$H MRI counterparts.

**Figure 5:** Comparison between the metabolic kinetics afforded by FT and by PRO-SLAM when targeting *in vivo* placentas (P1-P3) in a pregnant rat by hyperpolarized $^{13}$C$_1$-pyruvate injections. Spectral images corresponding to (a) the $^{13}$C pyruvate peak and (b) the $^{13}$C lactate peak, are displayed after reconstruction by either 2D FT (upper row in each panel) or PRO-SLAM (lower row in each panel) protocols. Compartments were drawn over the main placentas as shown on the enlarged right-hand $^1$H images.

**Figure 6:** Kinetic curves extracted from the different compartments and metabolites by PRO-SLAM and FT methods, as identified in the study introduced in Figure 5. Notice their similar time courses.

**Figure 7:** Comparison between the metabolic kinetics afforded by FT and PRO-SLAM for the $^{13}$C$_1$-alanine residue arising in the study introduced in Supporting Figure S1 as #2. Four compartments including two placentas (P1, P2), a fetal liver (L) and a maternal kidney (K) were defined for the PRO-SLAM processing (c). Results indicate that pyruvate is converted into alanine in the liver of the fetus already in the first scan, while peaks in the placentas peak for the second MRSI repetition. Kinetic curves (b) showing results based on FT and PRO-SLAM processings.

**Figure 8:** (a) *In vivo* transport and perfusion of hyperpolarized $^{13}$C urea, demonstrated in maternal organs of a pregnant rat. MRI images reconstructed with FT (top row) and PRO-SLAM (second row) reveal noticeable differences in the spatial resolution of the methods. The PRO-SLAM reconstruction focused on the three compartments shown on (c), entailing two maternal kidneys (K1, K2) and the maternal vena cava (VC). (b) Kinetic curves measuring $^{13}$C urea content across time in the targeted regions for both reconstruction methods.

**Supporting Information Figure S1**: Normalized magnitude profiles corresponding to the 1H MRSI scans in Fig. 2, extracted for the columns highlighted in that Figure with a dashed line. FT profiles are drawn in red, PRO-SLAM profiles in blue, and black lines were extracted from the high-definition image (Fig. 2e). The upper row corresponds to an acquired phase encoding matrix of 16x16, the middle row plots correspond to a 32x32 acquisition, and the bottom row to a 64x64 one. The left column plots were taken from the water spectral bin while the right column plots are taken from the acetone spectral bin.

**Supporting Information Figure S2**: Metabolism of $^{13}C_1$-enriched pyruvate arising from dDNP MRSI in five additional studies, complementing the results shown in Figure 4. Every panel study presents images corresponding to the lactate spectral bin and the pyruvate peak, with FT images on the top and PRO-SLAM is on the bottom row. The regions that were segmented and used as input for PRO-SLAM are highlighted on the accompanying $^1H$ anatomical images. Studies focused mainly on placentas (P), but also a maternal artery (A) and kidney (K) were occasionally defined as compartments for the PRO-SLAM regularization step. Study 6 was carried out at a slightly earlier (E16) pregnancy stage and hence the segmented ROIs target the full fetoplacental (FP) units.

**Supporting Information Figure S3:** Comparisons between *in vivo* metabolite kinetics based on hyperpolarized $^{13}C$ MRSI experiments reconstructed with PRO-SLAM against those reconstructed byFT, when focusing on the spectral peaks corresponding to $^{13}C_1$-pyruvate and $^{13}C_1$-lactate for the studies presented in Supporting Figure S1. Curves represent compartment means of signal intensities recovered for the various regions indicated in Studies 2-6 in Supporting Figure S1, averaging image values in over regions of interest that were predefined for PRO-SLAM. Notice how lactate curves demonstrate for both reconstruction methods an increase in placental lactate during the first 25 sec, followed by a steady decline.

**Supporting Information Figure S4**: Additional perfusion examples comparing PRO-SLAM and FT results arising upon injecting hyperpolarized $^{13}C$ urea on a pregnant rat. The various regions selected are highlighted in the accompanying $^1H$ MRI images, and the time-dependent signal intensities in the accompanying graphs.

**Supporting Information Table S1**: Deviation from the reference image ($\square_{ref}$) and edge contrast measurements done on proton chemical images from Fig. 2 for acquired *k*-matrix sizes of 16x16, 32x32 and 64x64.

**Supporting Information Table S2**: ROI means measured for various spectral bin images in the proton MRSI results in Fig. 2. The ROI's denoted as ROI1-4 were defined as four quadrants defined by bisecting a circle surrounding the water tube in four identical portions . All ROI's were detected automatically. Curve fitting with linear regression gave a relation of ROI_mean_P-S = 1.01*ROI_mean_FT with an R squared of 0.99.

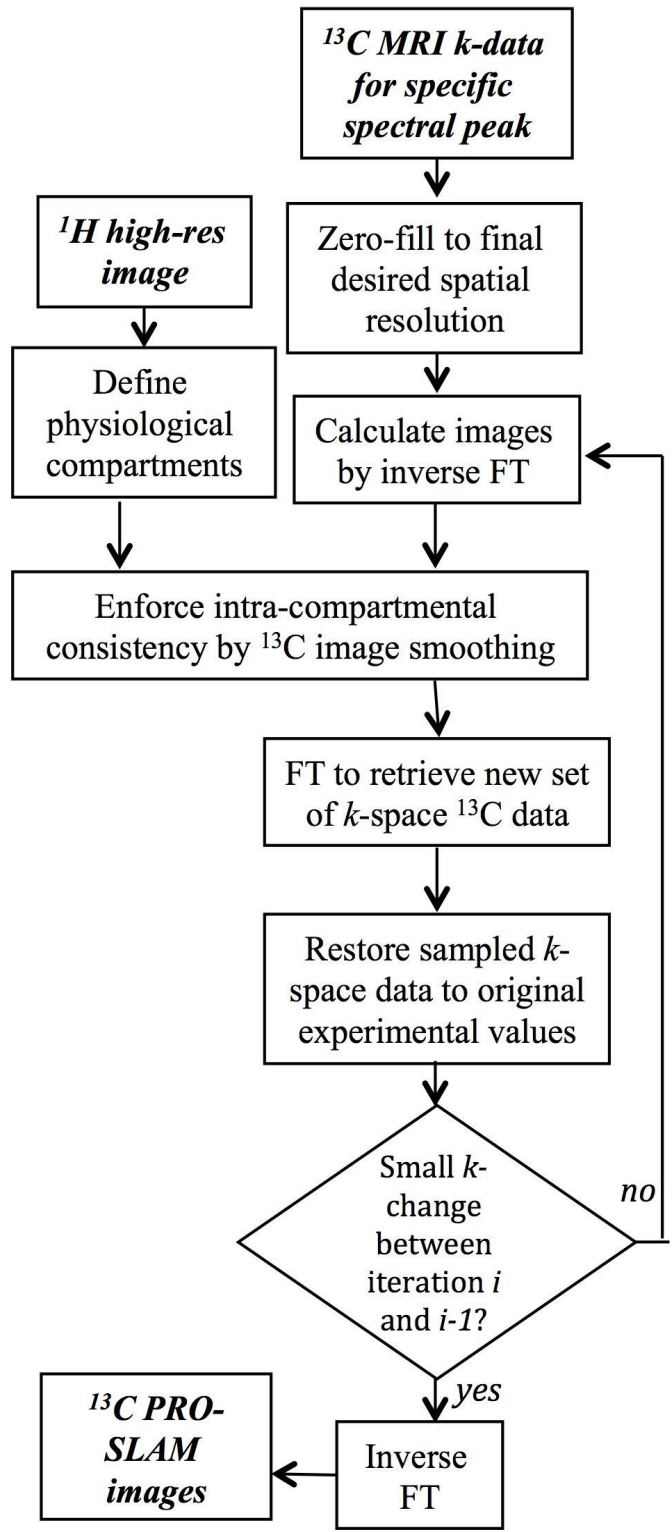

CSI measured on an enriched $^{13}$C glucose sample in thermal conditions

- Peak
- Noise
- $^{13}$C MRSI spectral dimension

**(a) FT k-processing**

**2D image - peak**     **2D image, noise RMS**

**(b) PRO-SLAM processing**

**(c) $SNR_{P-S}/SNR_{FT}$**     **(d) $^1$H reference image**

$^{13}$C glucose     $H_2O$

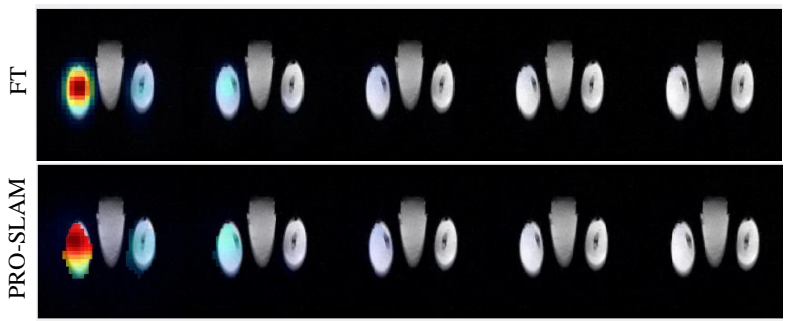

**(a) $^{13}C_1$ Pyruvate**

FT / PRO-SLAM

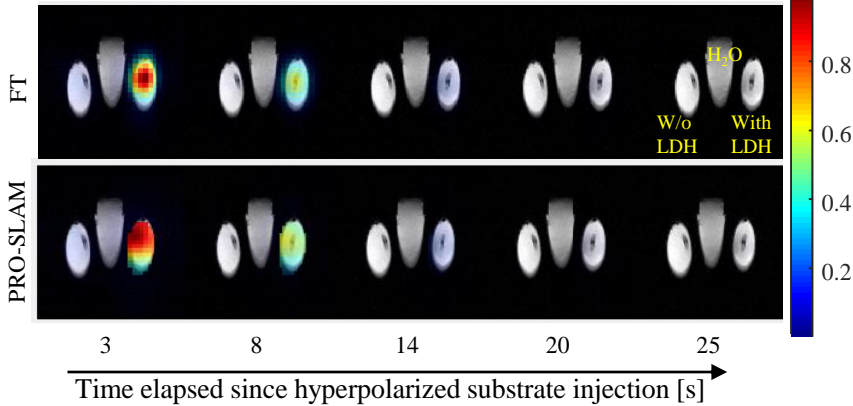

**(b) $^{13}C_1$ Lactate**

FT / PRO-SLAM

W/o LDH — H$_2$O — With LDH

3    8    14    20    25

Time elapsed since hyperpolarized substrate injection [s]

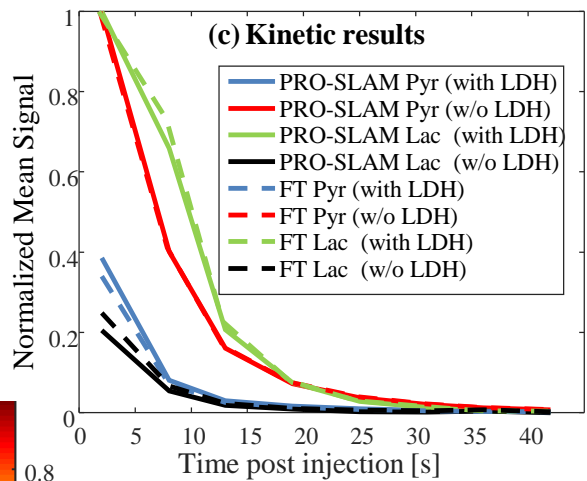

**(c) Kinetic results**

- PRO-SLAM Pyr (with LDH)
- PRO-SLAM Pyr (w/o LDH)
- PRO-SLAM Lac (with LDH)
- PRO-SLAM Lac (w/o LDH)
- FT Pyr (with LDH)
- FT Pyr (w/o LDH)
- FT Lac (with LDH)
- FT Lac (w/o LDH)

Normalized Mean Signal vs Time post injection [s]

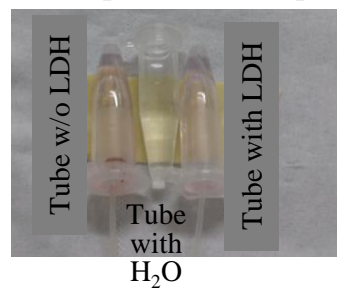

**(d) Experimental setup**

Tube w/o LDH | Tube with H$_2$O | Tube with LDH

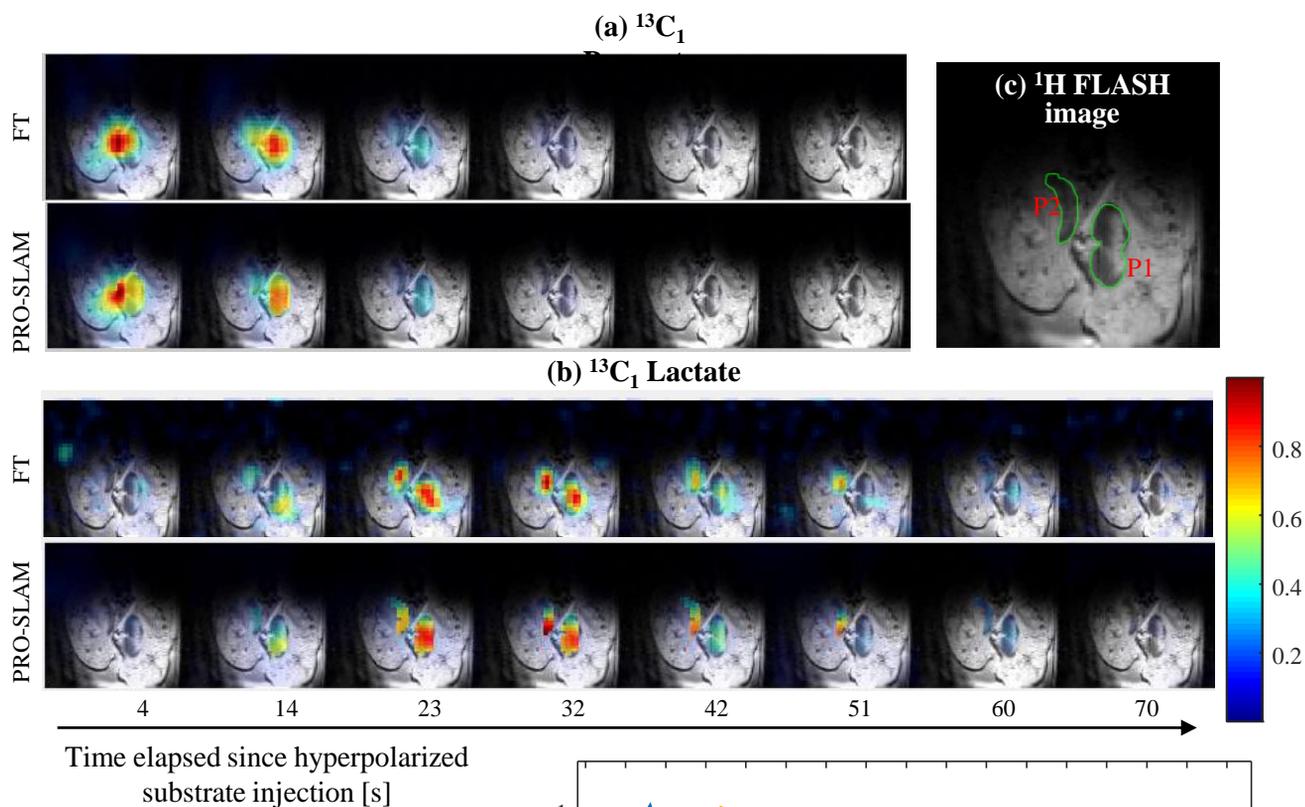

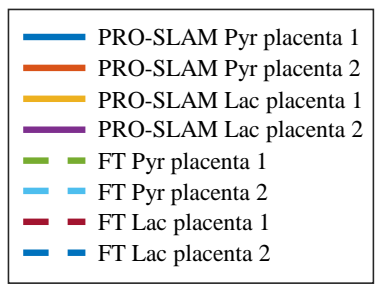

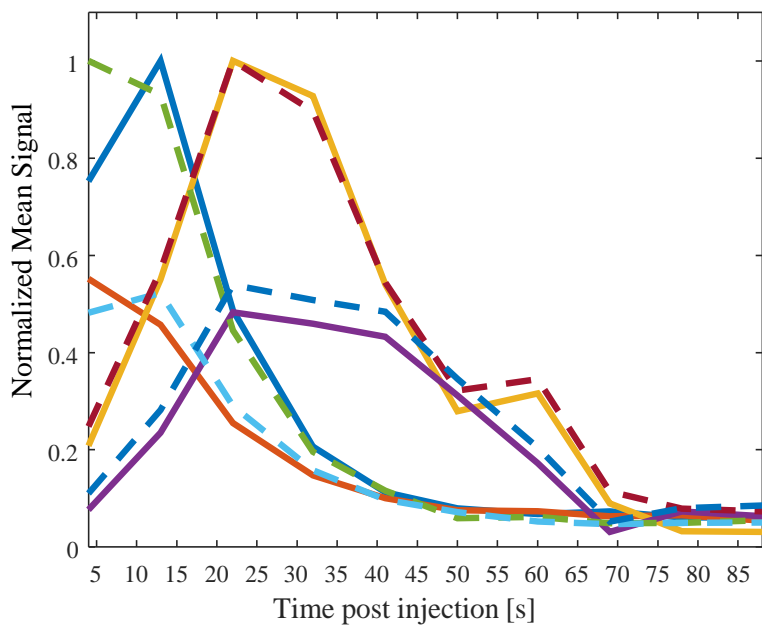

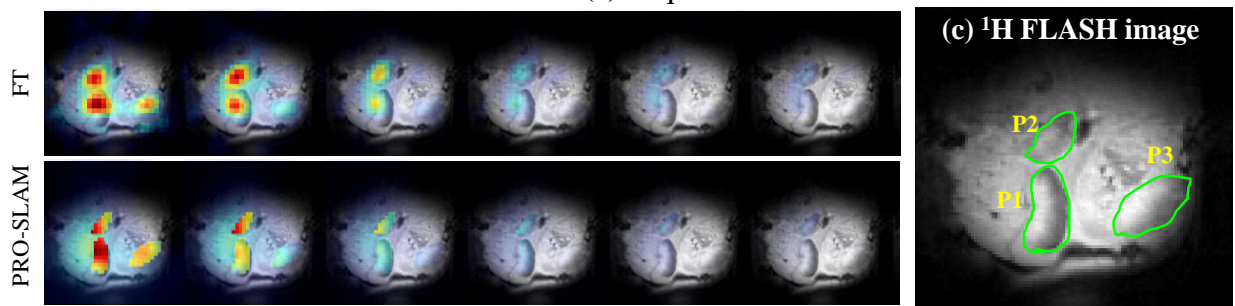

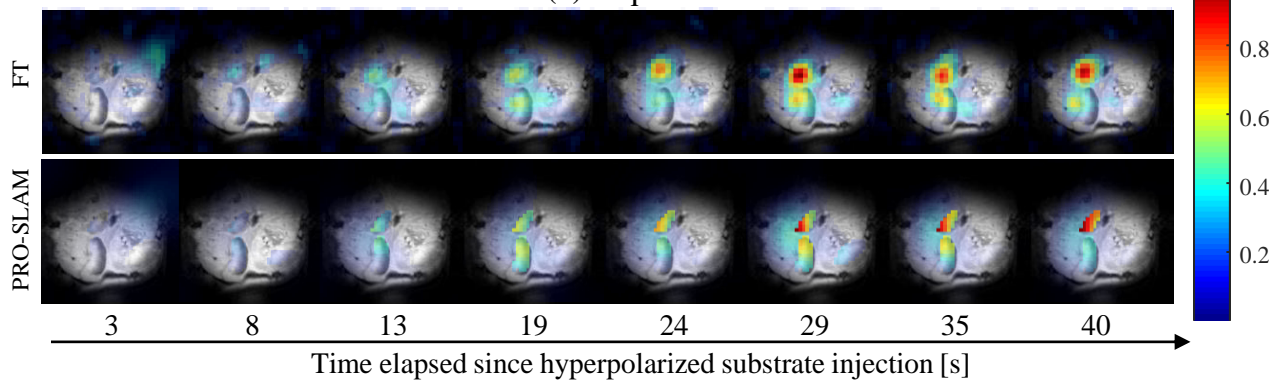

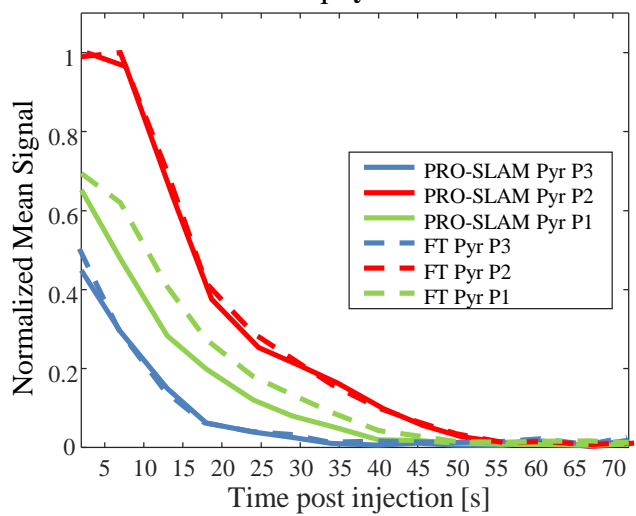
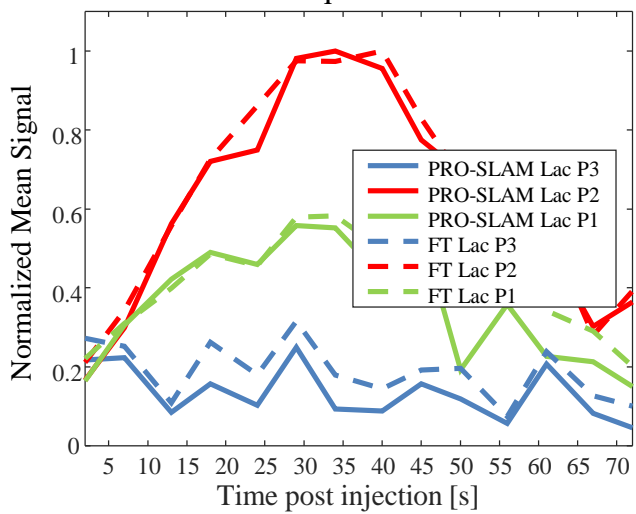

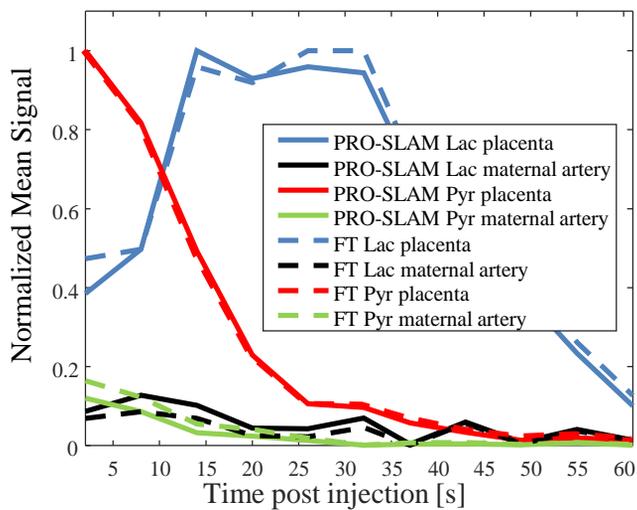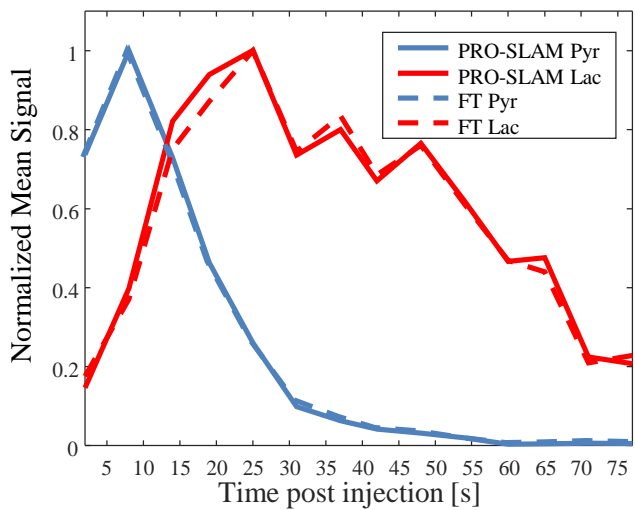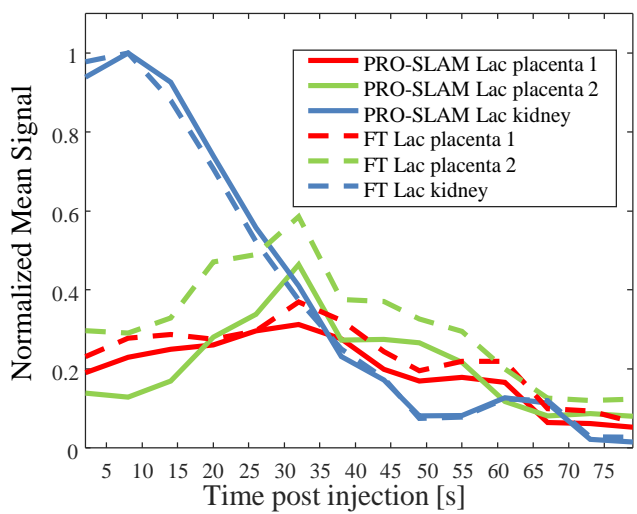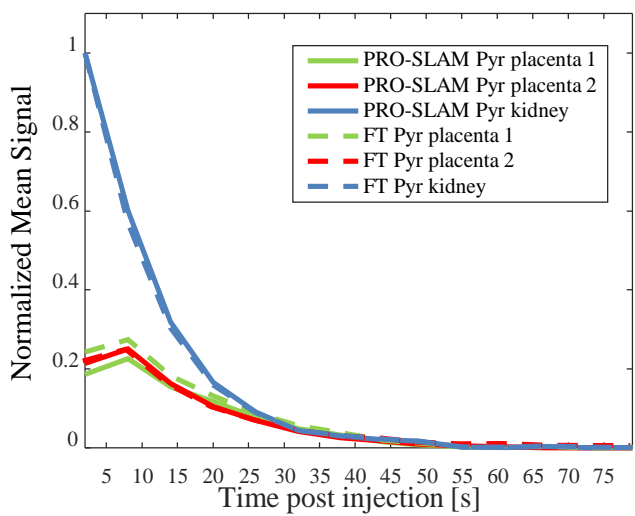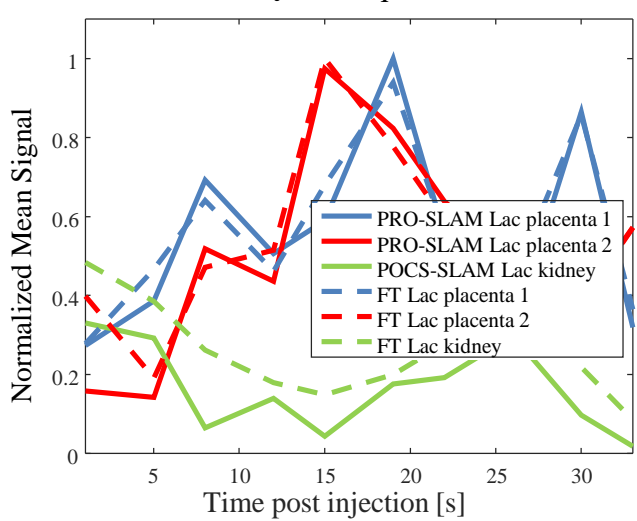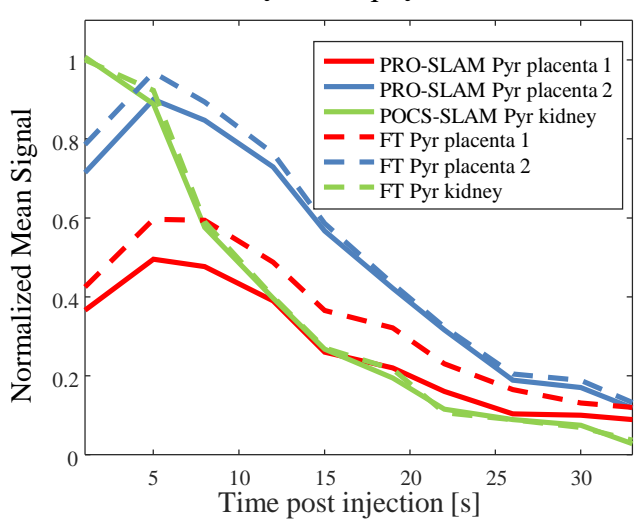

Supporting Information for

# Enhanced Hyperpolarized Chemical Shift Imaging Based on *a priori* Segmented Information

Gil Farkash, Stefan Markovic, Mihajlo Novakovic and Lucio Frydman*

*Department of Chemical and Biological Physics, Weizmann Institute of Science, Rehovot 76100, Israel*

1. **PRO-SLAM vs FT: I***n vivo* $^1$**H MRSI tests and quantifications**

While the manuscript's main emphasis is on $^{13}$C acquisitions, it is easier to quantify the performance of the PRO-SLAM approach on $^1$H MRSI experiments, as these are endowed with higher sensitivity and with the possibility of rerunning multiple tests under different acquisition but identical sample conditions. Representative examples of results arising from such MRSI acquisitions upon subjecting them to different processing modes are shown in Fig. 2 of the main text. Table S1 summarizes were calculated for a water region in Fig. 2a-2b and over the acetone tube ROI in Fig. 2c-2d. These measurements showed a significant improvement in the correlation of PRO-SLAM images to the reference image relative to FT for all the acquired $k$ matrix dimensions. These improvements, quantified in Table S1 as $\Delta_{ref} = 100 \cdot \frac{|s(x,y) - s_{ref}(x,y)|}{|s_{ref}(x,y)|}$, can also be appreciated in 1D profiles taken from these data, and displayed in Fig. S1. In general non-uniformities are still observed in the FT and PRO-SLAM results, as well as in the $^1$H image taken as reference; these are attributed mostly to B$_0$ and B$_1$ inhomogeneities.

Another important test for the PRO-SLAM reconstruction concerns its ability to leave the spatial edges of the resulting spectrally-resolved images, uncompromised. To explore this aspect the Δx contrast of the edges in the water/acetone phantom introduced Fig. S1, were measured after PRO-SLAM and FT reconstruction –in both cases after reaching a 128x128 matrix. Contrast for H$_2$O was quantified as $Contrast = 100 \cdot \left[\frac{S_{in} - S_{out}}{S_{in} + S_{out}}\right]$ where S$_{in}$ corresponds to the mean of 5 pixels within the water region and S$_{out}$ corresponds to the mean of 5 pixels outside the water region adjacent to the edge. For the acetone images, S$_{in}$ was averaged from 5 pixels inside the acetone region, while the S$_{out}$ averaged 5 pixels on an adjacent water region. These contrast values are also summarized in Table S1. For the water spectral bin, the contrast measurements confirmed the preservation of edge contrast in PRO-SLAM upon processing 32x32 and 64x64 matrices, and a contrast improvement when processing a 16x16 *k*-matrix. A similar trend was observed for the acetone spectral bin. These improvements can also be appreciated in the 1D profiles displayed in Fig. S1.

|  | **Acquired *k* Matrix** | **Deviation from Reference Image (%)** | **Edge Contrast (%)** |
|---|---|---|---|
| **Water image FT** | 16x16 | 5.5 | 77.8 |
|  | 32x32 | 9.8 | 91.2 |
|  | 64x64 | 4.7 | 94.4 |
| **Water image PRO-SLAM** | 16x16 | 2.6 | 87 |
|  | 32x32 | 4.5 | 92.1 |
|  | 64x64 | 3.7 | 94.7 |
| **Acetone image FT** | 16x16 | 19 | 67.1 |
|  | 32x32 | 11.5 | 84.4 |
|  | 64x64 | 7.4 | 84.6 |
| **Acetone image PRO-SLAM** | 16x16 | 6.7 | 71.6 |
|  | 32x32 | 5.8 | 85.6 |
|  | 64x64 | 7 | 82.5 |

**Table S1**: Deviation from the reference image ($\Delta_{ref}$) and edge contrast measurements done on proton chemical images from Fig. 2 for acquired *k*-matrix sizes of 16x16, 32x32 and 64x64.

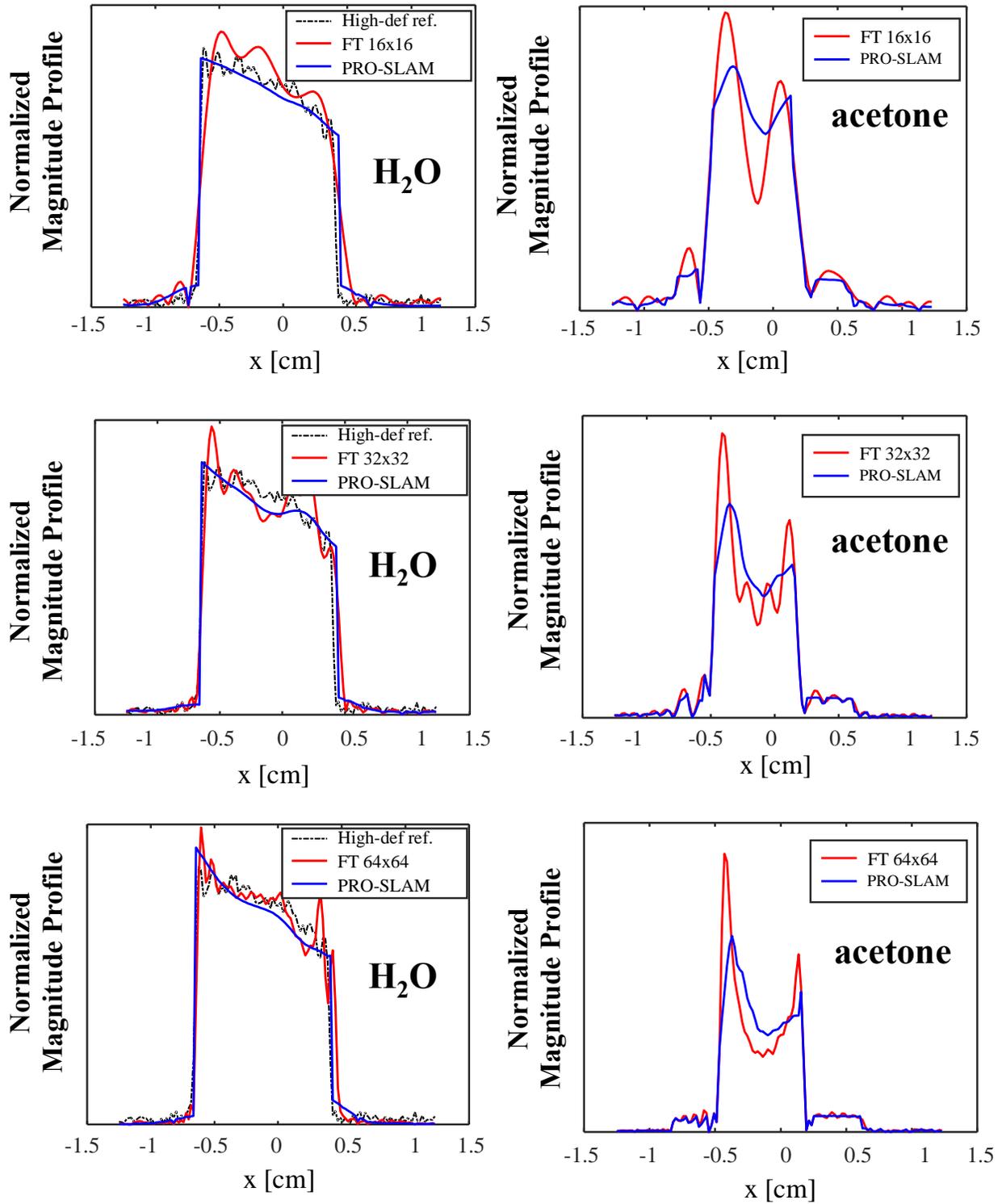

**Figure S1**: Normalized magnitude profiles corresponding to the 1H MRSI scans in Fig. 2, extracted for the columns highlighted in that Figure with a dashed line. FT profiles are drawn in red, PRO-SLAM profiles in blue, and black lines were extracted from the high-definition image (Fig. 2e). The upper row corresponds to an acquired phase encoding matrix of 16x16, the middle row plots correspond to a 32x32 acquisition, and the bottom row to a 64x64 one. The left column plots were taken from the water spectral bin while the right column plots are taken from the acetone spectral bin.

A final aspect deserving testing concerns the quantification reliability of the PRO-SLAM approach. The FT is a linear operation with a well known capacity to quantify concentrations according to spectral or image intensities; preserving such ability is an important demand for any image processing algorithm to be applied on an experiment that that needs to deliver reliable assessments of concentrations and kinetics. Table S2 analyses this aspect, by comparing average intensities arising for ROIs cut out from the four quadrants in the water and the acetone spectral images, upon processing the *k*-space data matrices by FT and by PRO-SLAM. It follows form these data that the intensity values provided by both methods are closely correlated; a linear regression of these means provides a proportionality between PRO-SLAM and FT intensities with a slope of 1.01 and an $r^2$-value of 0.99. This provides good additional validation regarding the linearity of both methods.

|  | PRO-SLAM 16x16 | FT 16x16 | PRO-SLAM 32x32 | FT 32x32 | PRO-SLAM 64x64 | FT 64x64 |
|---|---|---|---|---|---|---|
| **ROI1 mean – Acetone image** | 50 | 50 | 46 | 47 | 40 | 40 |
| **ROI2 mean – Acetone image** | 38 | 39 | 35 | 37 | 31 | 32 |
| **ROI3 mean – Acetone image** | 28 | 29 | 27 | 27 | 25 | 26 |
| **ROI4 mean – Acetone image** | 22 | 22 | 20 | 21 | 19 | 20 |
| **ROI1 mean – water image** | 140 | 133 | 134 | 131 | 122 | 119 |
| **ROI2 mean – water image** | 160 | 157 | 147 | 147 | 137 | 137 |
| **ROI3 mean – water image** | 206 | 203 | 199 | 198 | 188 | 189 |
| **ROI4 mean – water image** | 156 | 154 | 148 | 146 | 133 | 132 |

**Table S2**: ROI means measured for various spectral bin images in the proton MRSI results in Fig. 2. The ROI's denoted as ROI1-4 were defined as four quadrants defined by bisecting a circle surrounding the water tube in four identical portions . All ROI's were detected automatically. Curve fitting with linear regression gave a relation of ROI_mean_P-S = 1.01*ROI_mean_FT with an R squared of 0.99.

## 2. PRO-SLAM vs FT: Five additional pyruvate and two additional urea *in vivo* MRSI studies on pregnant rats

Figures 5 and 6 compared hyperpolarized *in vivo* MRSI experiments that monitored pyruvate enzymatic turnover to lactate in pregnant rats, processed by regular FT vs PRO-SLAM. Figure S2 summarizes five additional studies focusing on both maternal and fetoplacental metabolism for different animals, with raw data taken from the study in Ref. 13. Figure S3 compares the pyruvate and lactate kinetics afforded by the FT and PRO-SLAM processing for these five additional studies. In general, good agreement is observed between the two methods.

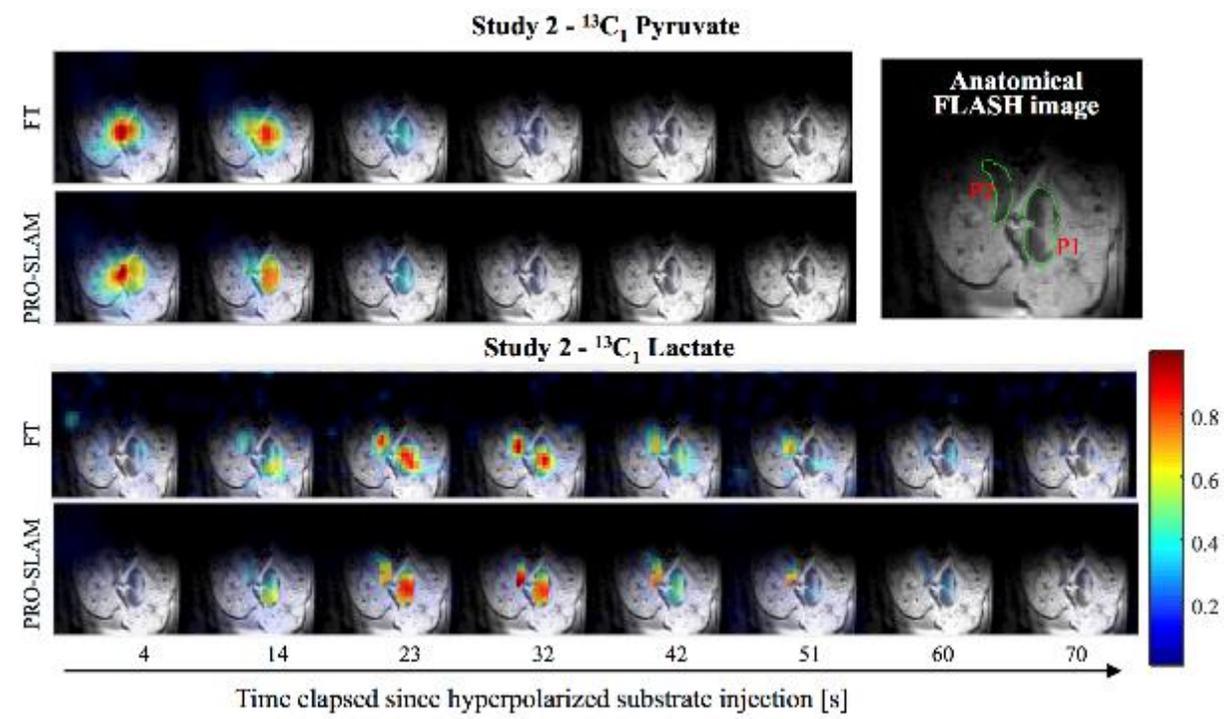
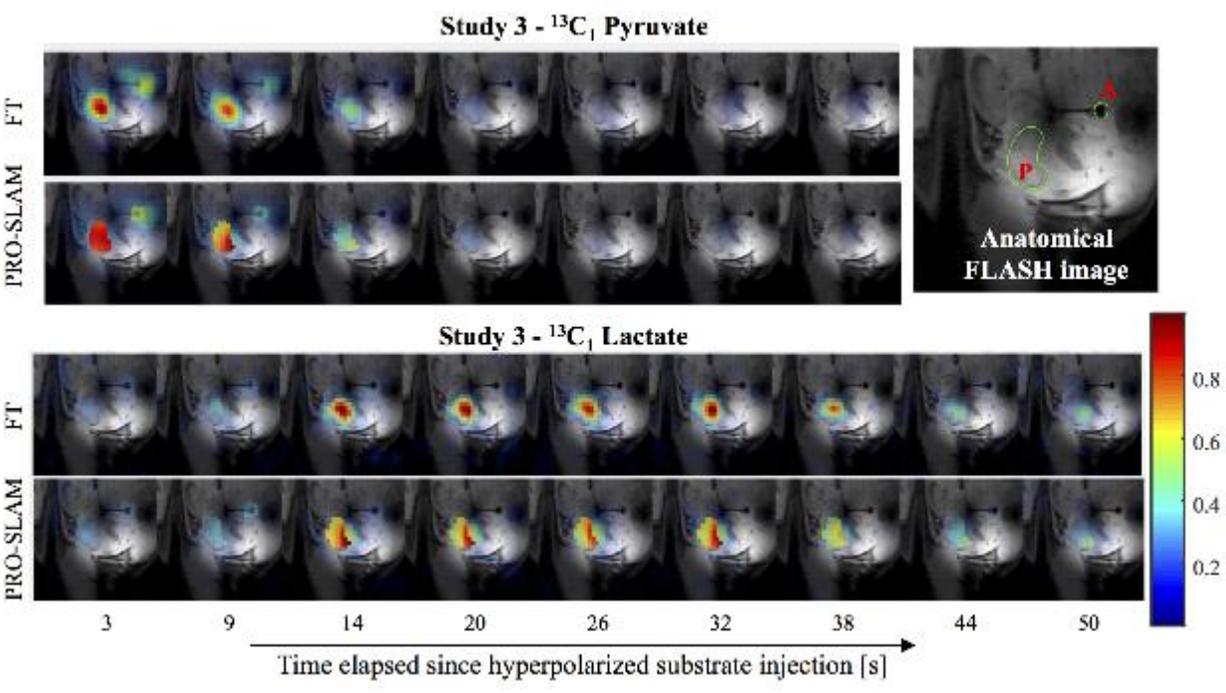

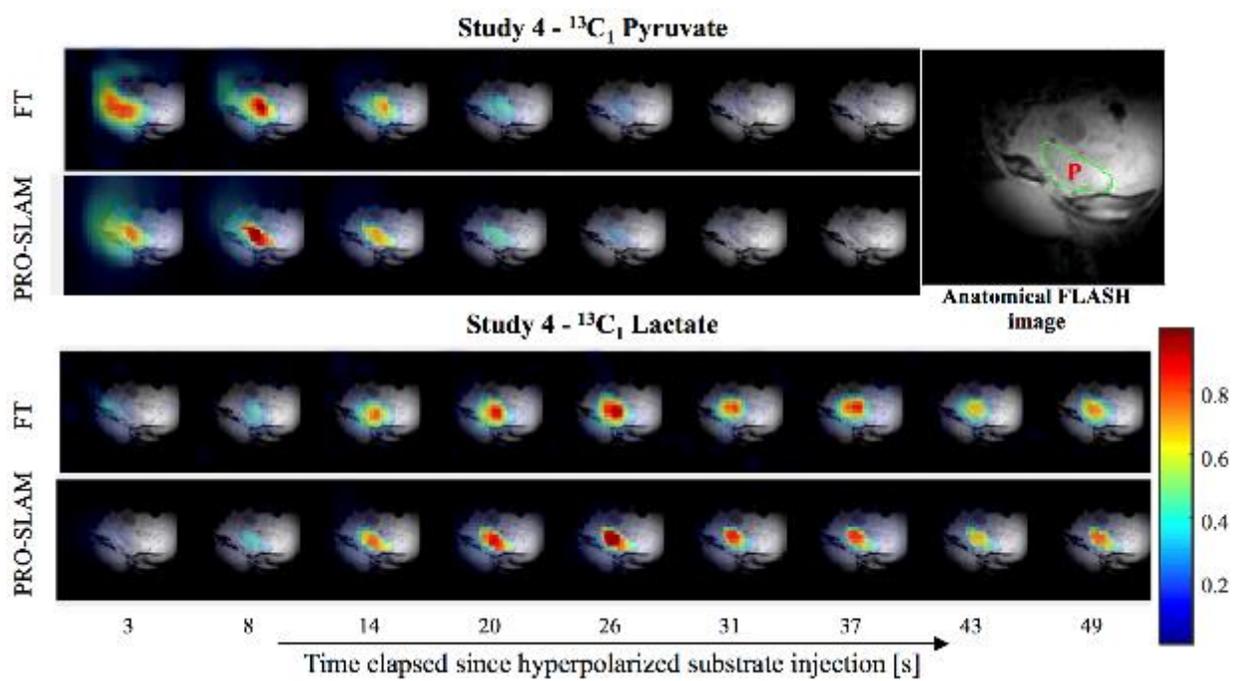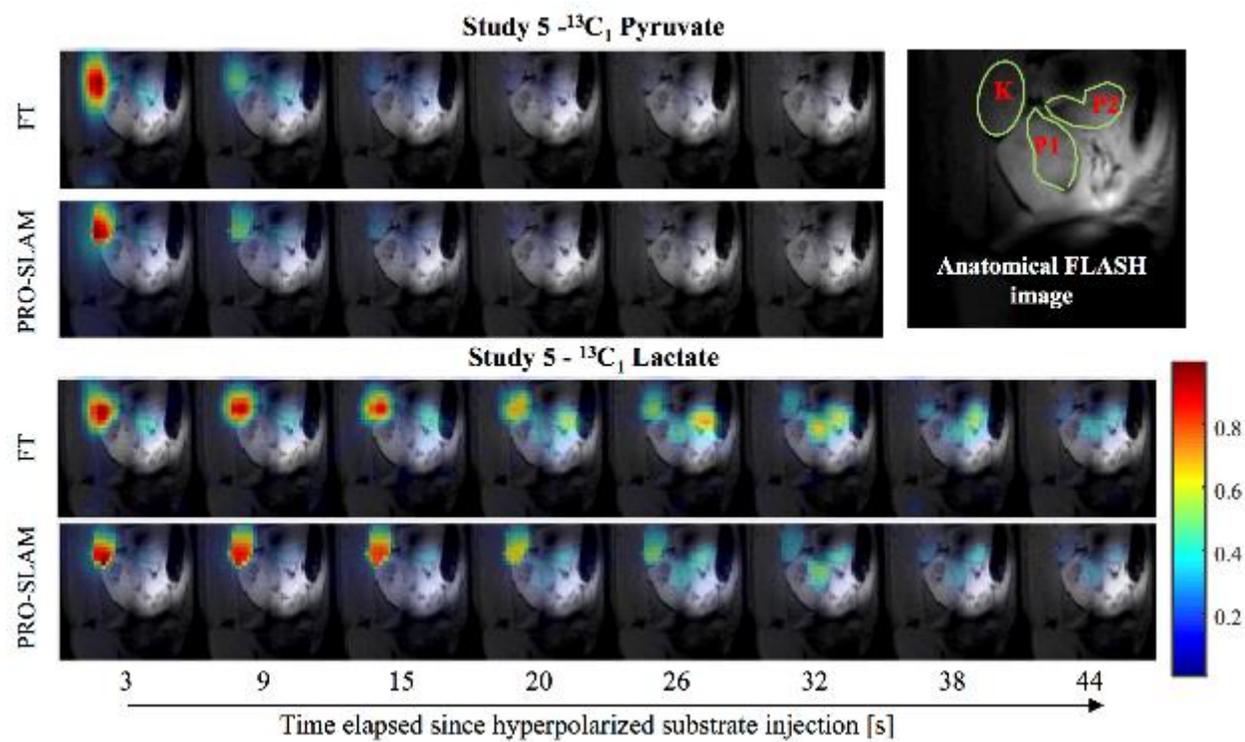

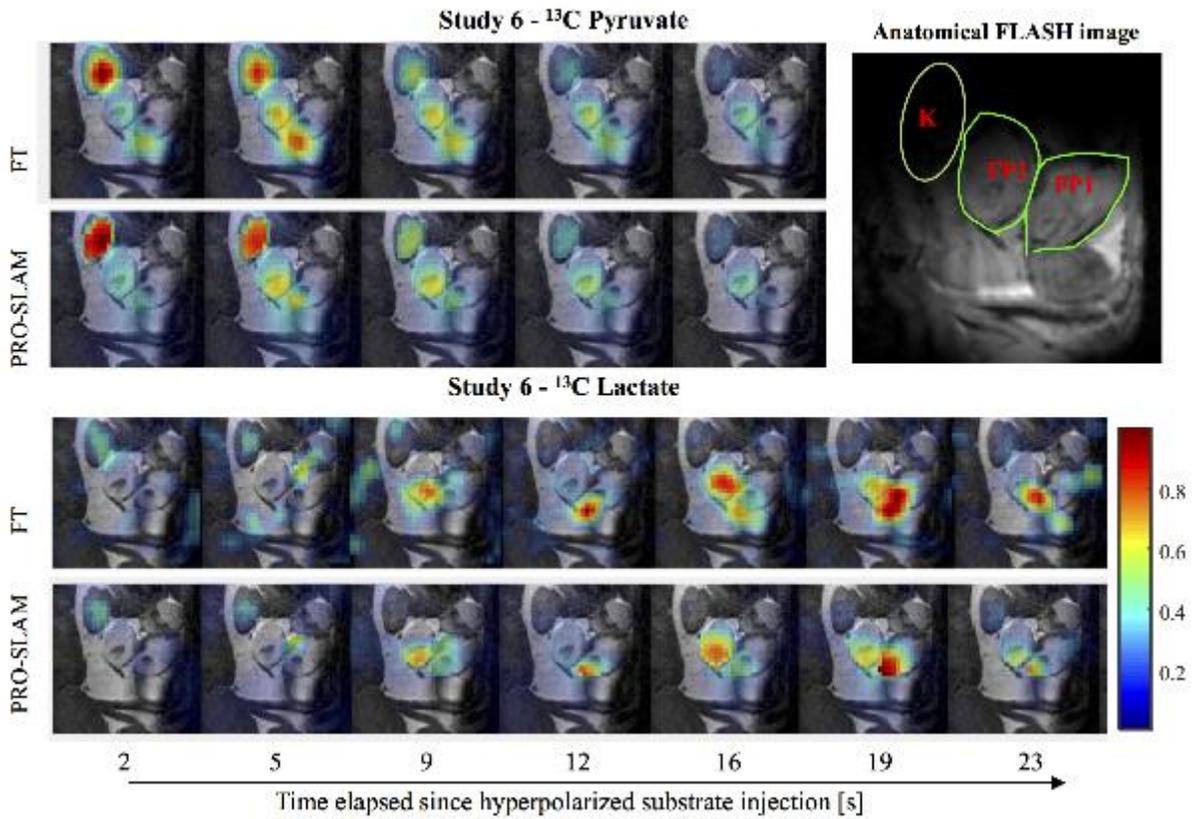

**Figure S2**: Metabolism of $^{13}C_1$-enriched pyruvate arising from dDNP MRSI in five additional studies, complementing the results shown in Figure 4. Every panel study presents images corresponding to the lactate spectral bin and the pyruvate peak, with FT images on the top and PRO-SLAM is on the bottom row. The regions that were segmented and used as input for PRO-SLAM are highlighted on the accompanying $^1H$ anatomical images. Studies focused mainly on placentas (P), but also a maternal artery (A) and kidney (K) were occasionally defined as compartments for the PRO-SLAM regularization step. Study 6 was carried out at a slightly earlier (E16) pregnancy stage and hence the segmented ROIs target the full fetoplacental (FP) units.

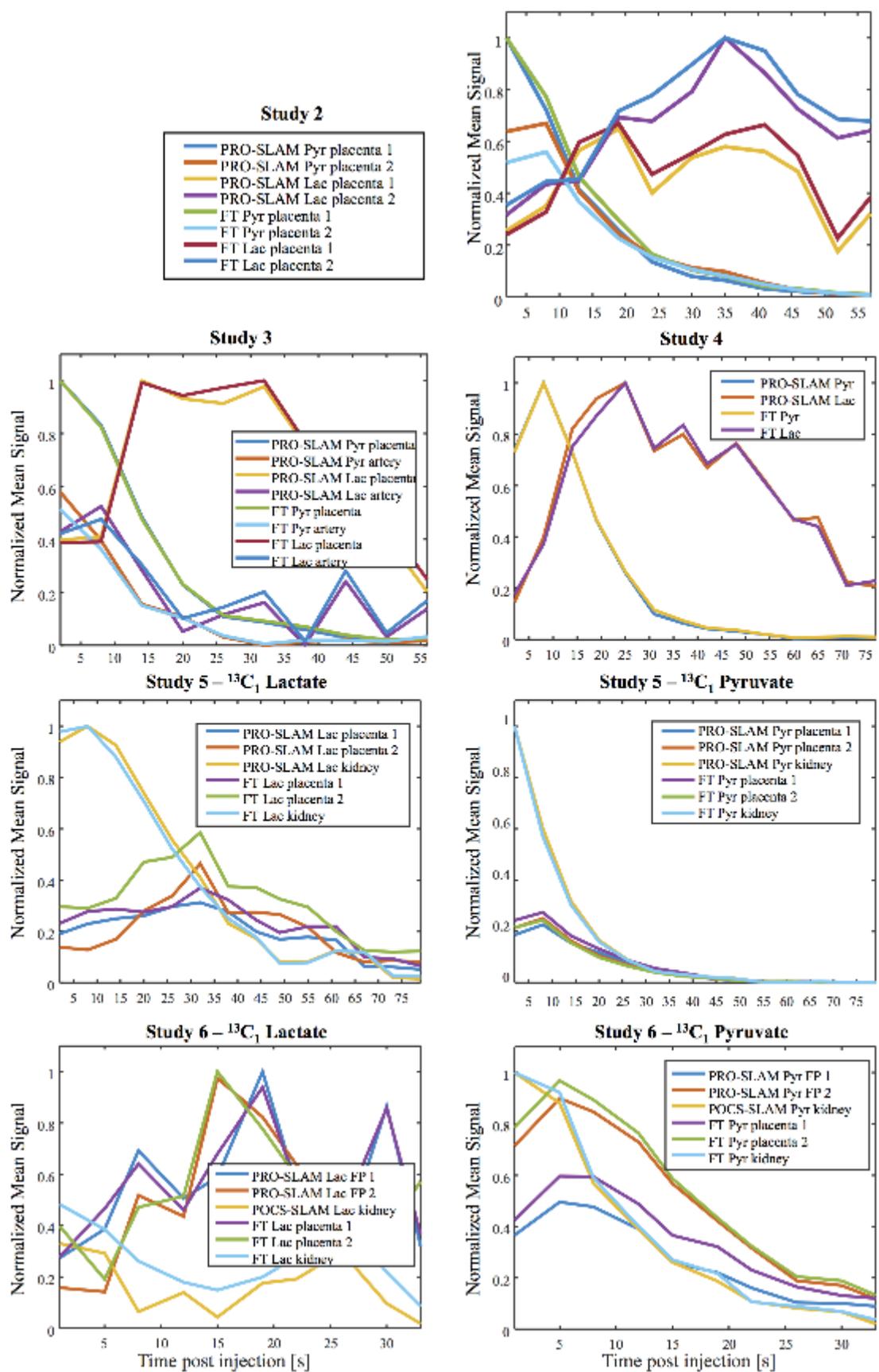

**Figure S3:** Comparisons between *in vivo* metabolite kinetics based on hyperpolarized $^{13}$C MRSI experiments reconstructed with PRO-SLAM against those reconstructed by FT, when focusing on the spectral peaks corresponding to $^{13}C_1$-pyruvate and $^{13}C_1$-lactate for the studies presented in Figure S1. Curves represent compartment means of signal intensities recovered for the various regions indicated in Studies 2-6 in Figure S1, averaging image values in over regions of interest that were predefined for PRO-SLAM. Notice how lactate curves demonstrate for both reconstruction methods an increase in placental lactate during the first 25 sec, followed by a steady decline.

Figure 8 focused on transport of hyperpolarized urea into maternal compartments. Figure S4 presents two additional urea studies, arising once again from raw data presented in Ref. 23, focusing on fetoplacental compartments. Notice the better localization and sensitivity improvements provided in these cases by the PRO-SLAM processing.

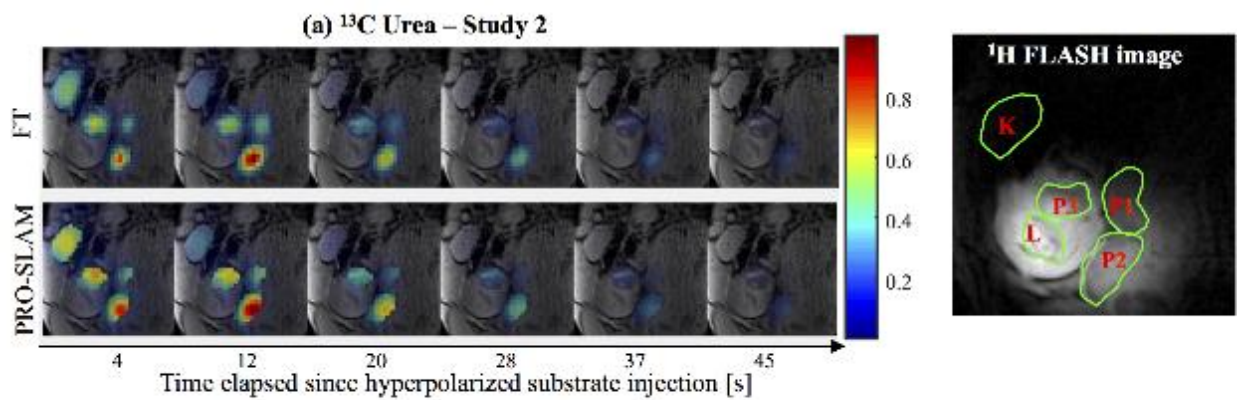

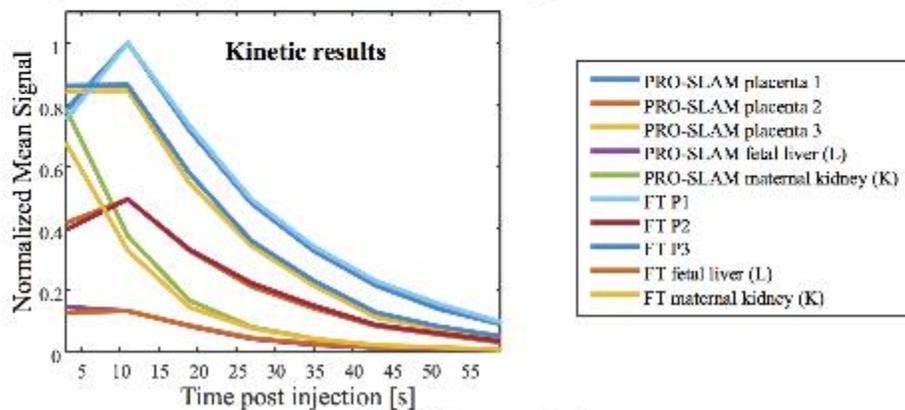

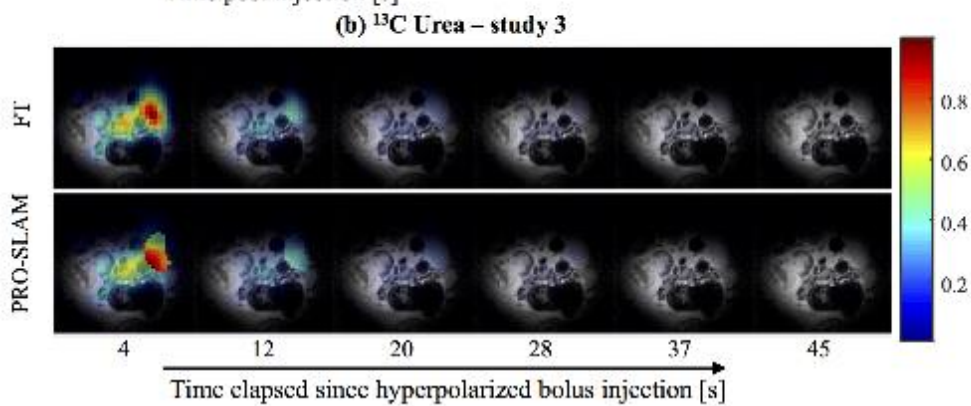

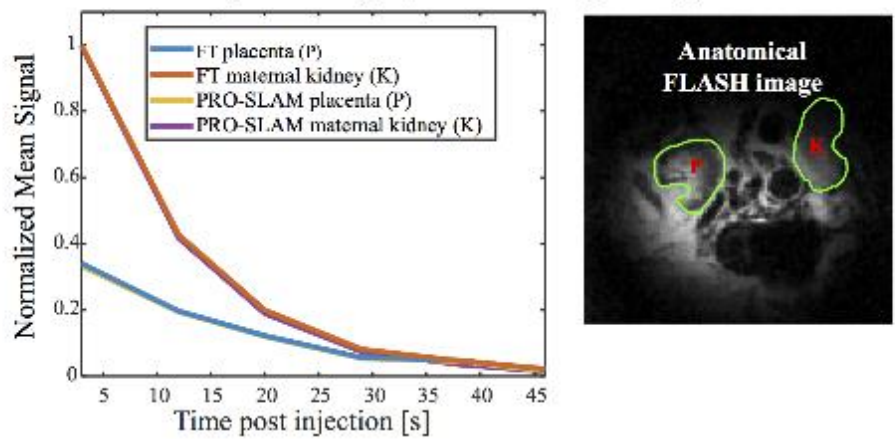

**Figure S4**: Additional perfusion examples comparing PRO-SLAM and FT results arising upon injecting hyperpolarized $^{13}$C urea on a pregnant rat. The various regions selected are highlighted in the accompanying $^1$H MRI images, and the time-dependent signal intensities in the accompanying graphs.